\documentclass[letter]{aa} 
\usepackage{graphicx}   
\usepackage{natbib}   
\usepackage{color}   
\bibpunct{(}{)}{;}{a}{}{,} % to follow A&A style   

\newcommand{\micron}{\mbox{$\mu$m}}

\newcommand{\HII}{\ion{H}{ii}}   
\newcommand{\HI}{\ion{H}{i}}   
   
\newcommand{\NII}{[\ion{N}{ii}]}

\def\farcs{\hbox{$.\!\!^{\prime\prime}$}}
\usepackage[colorlinks]{hyperref}
\begin{document}
\title{Properties of compact 250\,\micron\ emission and \HII\ regions in M\,33 (\texttt{HERM33ES})\thanks{\textit{Herschel} is an ESA space observatory with science instruments provided by European-led Principal Investigator consortia and with important participation from NASA.}}
   \author{
     S.\,Verley\inst{1} \and
     M.\,Rela\~{n}o\inst{2} \and
     C.\,Kramer\inst{3} \and
     E.\,M.\,Xilouris\inst{4} \and
     M.\,Boquien\inst{5} \and
     D.\,Calzetti\inst{5} \and 
     F.\,Combes\inst{6} \and
     C.\,Buchbender\inst{3}  \and
     J.\,Braine\inst{7} \and 
     G.\,Quintana-Lacaci\inst{3}  \and
     F.\,S.\,Tabatabaei\inst{8} \and
     S.\,Lord\inst{9} \and
     F.\,Israel\inst{10} \and
     G.\,Stacey\inst{11} \and
     P.\,van der Werf\inst{10}
          }   

   \institute{
     Dept. F\'{i}sica Te\'{o}rica y del Cosmos, Universidad de Granada, Spain -- \email{simon@ugr.es}
     \and %no15:
     Institute of Astronomy, University of Cambridge, Madingley Road, 
     Cambridge CB3 0HA, England
     \and %no25:
     Instituto Radioastronomia Milimetrica (IRAM), 
     Av. Divina Pastora 7, Nucleo Central, E-18012 Granada, Spain
     \and %no21:
     Institute of Astronomy and Astrophysics, National Observatory of Athens, 
     P. Penteli, 15236 Athens, Greece
     \and %no5
     Department of Astronomy, University of Massachusetts, Amherst, 
     MA 01003, USA 
     \and %no7:
     Observatoire de Paris, LERMA, 61 Av. de l'Observatoire, 
     75014 Paris, France
     \and %no6
     Laboratoire d'Astrophysique de Bordeaux, Universit\'{e} Bordeaux 1, 
     Observatoire de Bordeaux, OASU, UMR 5804, CNRS/INSU, B.P. 89, 
     Floirac F-33270, France
     \and %no4
     Max-Planck-Institut f\"ur Radioastronomie (MPIfR), 
     Auf dem H\"ugel 69, D-53121 Bonn, Germany
     \and %no12:
     Infrared Processing and Analysis Center, MS 100-22
     California Institute of Technology, Pasadena, CA 91125, USA
     \and %no10:
     Leiden Observatory, Leiden University, PO Box 9513, NL 2300 RA Leiden, The Netherlands
     \and %no20:
     Department of Astronomy, Cornell University, Ithaca, NY 14853, USA
}

%   \offprints{S. Verley, \email{simon@ugr.es}}   
%   \date{Received / Accepted }   
      
% context, aims, methods, results, conclusions  
   \abstract
% Context:
{}
% Aims:   
   {Within the framework of the \texttt{HERM33ES} Key Project, using the high resolution and sensitivity of the \textit{Herschel} photometric data, we study the compact emission in the Local Group spiral galaxy M\,33 to investigate the nature of the compact SPIRE emission sources. We extracted a catalogue of sources at 250\,\micron\ in order to investigate the nature of this compact emission. Taking advantage of the unprecedented \textit{Herschel} resolution at these wavelengths, we also focus on a more precise study of some striking H$\alpha$ shells in the northern part of the galaxy.}
% Methods:
   {We present a catalogue of 159 compact emission sources in M\,33 identified by SExtractor in the 250\,\micron\ SPIRE band that is the one that provides the best spatial resolution. We also measured fluxes at 24\,\micron\ and H$\alpha$ for those 159 extracted sources. The morphological study of the shells also benefits from a multiwavelength approach including H$\alpha$, far-ultraviolet from \textit{GALEX}, and infrared from both \textit{Spitzer} IRAC 8\,\micron\ and MIPS 24\,\micron\ in order to make comparisons.}
% Results:
   {For the 159 compact sources selected at 250\,\micron, we find a very strong Pearson correlation coefficient with the MIPS 24\,\micron\ emission ($r_{24} = 0.94$) and a rather strong correlation with the H$\alpha$ emission, although with more scatter ($r_{\rm H\alpha} = 0.83$). The morphological study of the H$\alpha$ shells shows a displacement between far-ultraviolet, H$\alpha$, and the SPIRE bands. The cool dust emission from SPIRE clearly delineates the H$\alpha$ shell structures.}
% Conclusions:
   {The very strong link between the 250\,\micron\ compact emission and the 24\,\micron\ and H$\alpha$ emissions, by recovering the star formation rate from standard recipes for \HII\ regions, allows us to provide star formation rate calibrations based on the 250\,\micron\ compact emission alone. The different locations of the H$\alpha$ and far-ultraviolet emissions with respect to the SPIRE cool dust emission leads to a dynamical age of a few Myr for the H$\alpha$ shells and the associated cool dust.}
   \authorrunning{Verley et al.} 
  \titlerunning{Properties of compact 250\,\micron\ emission and \HII\ regions in M\,33 (\texttt{HERM33ES})}   
   \maketitle   
%________________________________________________________________

   \section{Introduction} %%%%%%%%%%%%%%%%%%%%%%%%%%%%%%%%%%%%%%%%%%%
Within the framework of the open time key project ``{\it Herschel} M\,33 extended survey ({\tt HERM33ES})'', we are studying the galaxy M\,33 to understand the origin of various diagnostic lines in heating and cooling and other processes in the interstellar medium (ISM). We refer the reader to \citet{2010A&A...501L...1K} for more details about the overall {\tt HERM33ES} project goals, as well as for a first presentation of the PACS and SPIRE maps of the entire galaxy, together with spatially averaged spectral energy distributions.

The local group, late-type, spiral galaxy --M\,33-- is an ideal target for studying the detailed processes of star formation (SF). Indeed, the distance, 840\,kpc \citep{1991ApJ...372..455F}, and inclination, 56 degrees \citep{1994ApJ...434..536R}, of M\,33 allow us to reach an unprecedented spatial resolution in the far-infrared of an external spiral galaxy and investigate the phenomena in the interstellar medium which lead to the SF. With {\it Herschel} \citep{2010A&A...001L...1P}, we are able to resolve 85\,pc, 109\,pc, and 170\,pc structures of the cool dust in the SPIRE 250, 350, and 500\,\micron\ bands, respectively, which is the typical spatial scale of giant \HII\ regions and their complexes. In this article, we follow a two-fold approach in order to study the nature of the compact 250\,\micron\ emission and take advantage of the resolution and sensitivity reached by {\it Herschel} to resolve the cool dust emission associated with \HII\ regions.

\section{Data} \label{sec:obs}

\subsection{{\it Herschel} photometric data}

During the photometric observations of the {\tt HERM33ES} key project, we mapped M\,33 with PACS and SPIRE instruments in parallel mode in two orthogonal directions, integrating 6.3\,hours (two AORs). In this letter, we particularly focus on the nature of the compact emission in the SPIRE bands. The reduction includes all basic pipeline steps plus a robust polynomial fit for the baseline, which improves the images by reducing scan line structures \citep[more details about the SPIRE observations are given in][]{2010A&A...501L...1K}. The r.m.s. noise levels of the SPIRE maps are 14.1, 9.2, and 8\,mJy/beam, at 250, 350, 500\,\micron. The mean measured spatial resolutions for Gaussian beams are $19\farcs7 \pm 1\farcs9$, $27\farcs0 \pm 4\farcs1$, $37\farcs2 \pm 2\farcs3$ for 250, 350, and 500\,\micron, respectively. The overall absolute calibration accuracy is estimated as within $\pm$15\%, following the SPIRE instrument paper \citep{2010A&A.....1L...1G}.

\begin{figure}
  \includegraphics[width=\columnwidth]{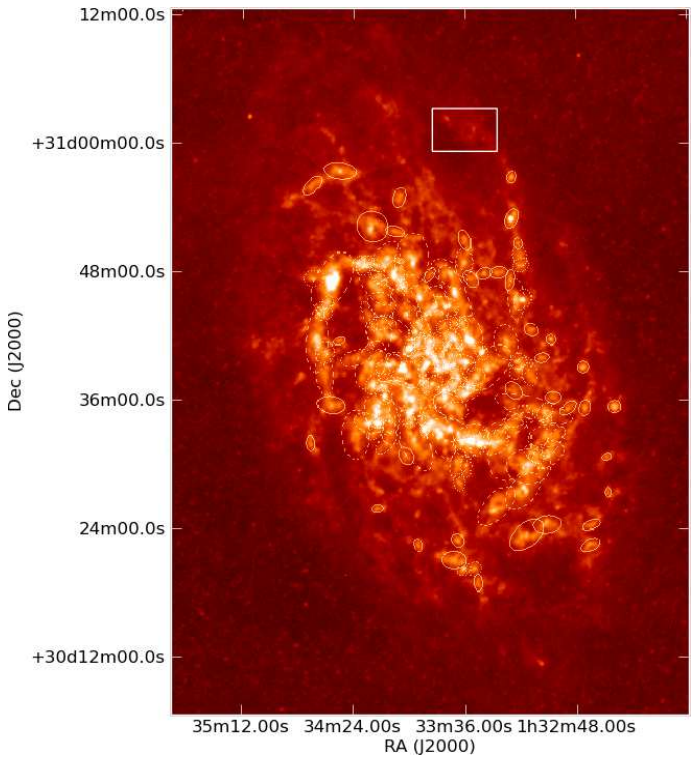}
\caption{SPIRE maps of M33 at 250\,$\mu$m (logarithmic scale). The catalogue of extracted sources is superimposed. The ellipses are indicative radii and are meant to show the position of the sources, they are more extended than where the photometry is actually performed. The white rectangle delimits the zone detailed in Fig.~\ref{fig:shell}.}
\label{fig:spire250cat}
\end{figure}

\subsection{Ancillary multiwavelength data}

To perform comparisons with the {\it Herschel} far-infrared emissions, we use the following multi-wavelength ancillary data: H$\alpha$ narrow-line image \citep{2000ApJ...541..597H} to trace the {\it current} unobscured SF ($<$10\,Myr), {\it GALEX} far-ultraviolet \citep{2007ApJS..173..185G} to trace the {\it recent} unobscured SF ($<$100\,Myr), and {\it Spitzer} IRAC 8\,\micron\ and MIPS 24\,\micron\ \citep{2007A&A...476.1161V,2009A&A...493..453V,2010A&A...510A..64V} infrared data to trace the obscured (probably recent) SF. In the H$\alpha$ map, the \NII\ contamination is less than 5\%, and we have corrected the flux for the foreground Galactic extinction, $A_{\rm H\alpha} = 0.17$ \citep{2009ApJ...699.1125R}. We consider typical uncertainties to be better than 15\% for the MIPS 24\,\micron\ \citep{2009A&A...493..453V} and the H$\alpha$ photometric measurements \citep{2000ApJ...541..597H}.

\vspace{-.5cm}
\section{Nature of the compact SPIRE 250\,\micron\ emission} \label{sec:sta}

In this section, we investigate the nature of the compact SPIRE 250\,\micron\ emission, in a statistical manner. To create a catalogue of compact emission sources in the SPIRE 250\,\micron\ band, we use the SExtractor software \citep{1996A&AS..117..393B}. The photometry of the sources is computed using the parameter {\sc flux\_iso} given by SExtractor, which uses isophotal photometry (sum of all the pixels above a threshold given by the lowest isophot: 16 times the background r.m.s.). Of course, the number, position, size, and photometry of the obtained sources can vary widely by choosing other input parameters in order to pursue a given scientific goal, but the above parameters are well-suited to recover as much of the compact 250\,\micron\ emission regions as we can, and lead to a final catalogue of 159 sources. The catalogue of extracted sources is overplotted in Fig.~\ref{fig:spire250cat}.

Interestingly, we detect almost no sources in the inter-arm regions. Since most of the extracted objects are along the spiral pattern of the galaxy, we believe many of the sources may be directly linked to SF. This suggests that the SPIRE 250\,\micron\ compact emission could be a reliable SF tracer in the vicinity of \HII\ regions. Special care must be taken for the most diffuse part of the 250\,\micron\ emission, which may be powered by a more general interstellar radiation field and may not be directly linked to current or recent SF \citep[see for instance][]{1996A&A...308..723I}. Detailed studies of the reliability of the diffuse PACS and SPIRE emission as SF tracers will be presented and discussed in related papers \citep[][Boquien et al., Calzetti et al., in prep.]{2010A&A...501L...1B}. For the present letter, we concentrate our preliminary work on the 250\,\micron\ compact emission and compare its properties with standard SF tracers such as the H$\alpha$ emission line and the 24\,\micron\ compact emission linked to \HII\ regions \citep{2005ApJ...633..871C,2007ApJ...666..870C}.

\begin{figure} [ht!]
\vspace{-.3cm}
\includegraphics[width=\columnwidth]{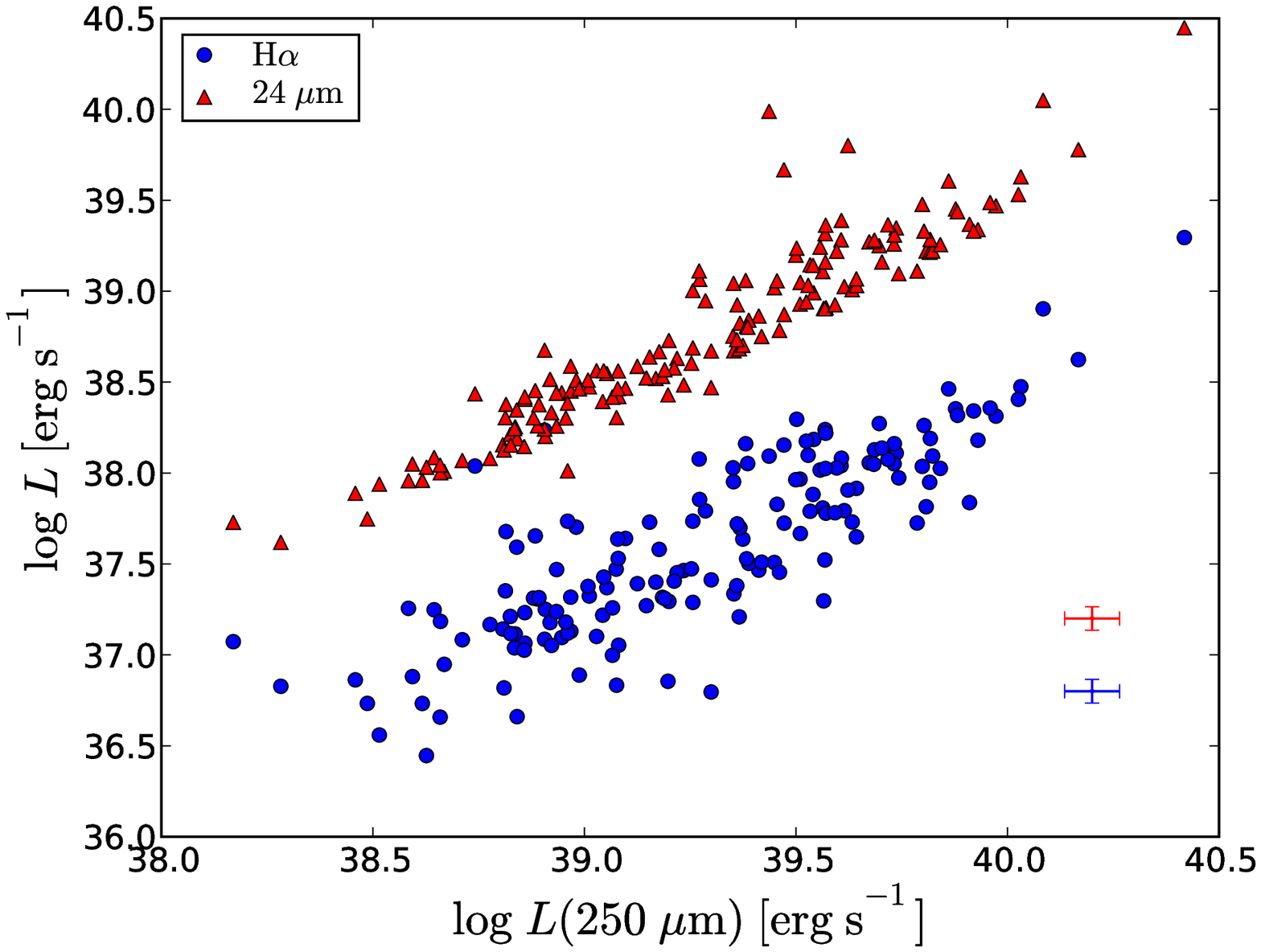}
\includegraphics[width=\columnwidth]{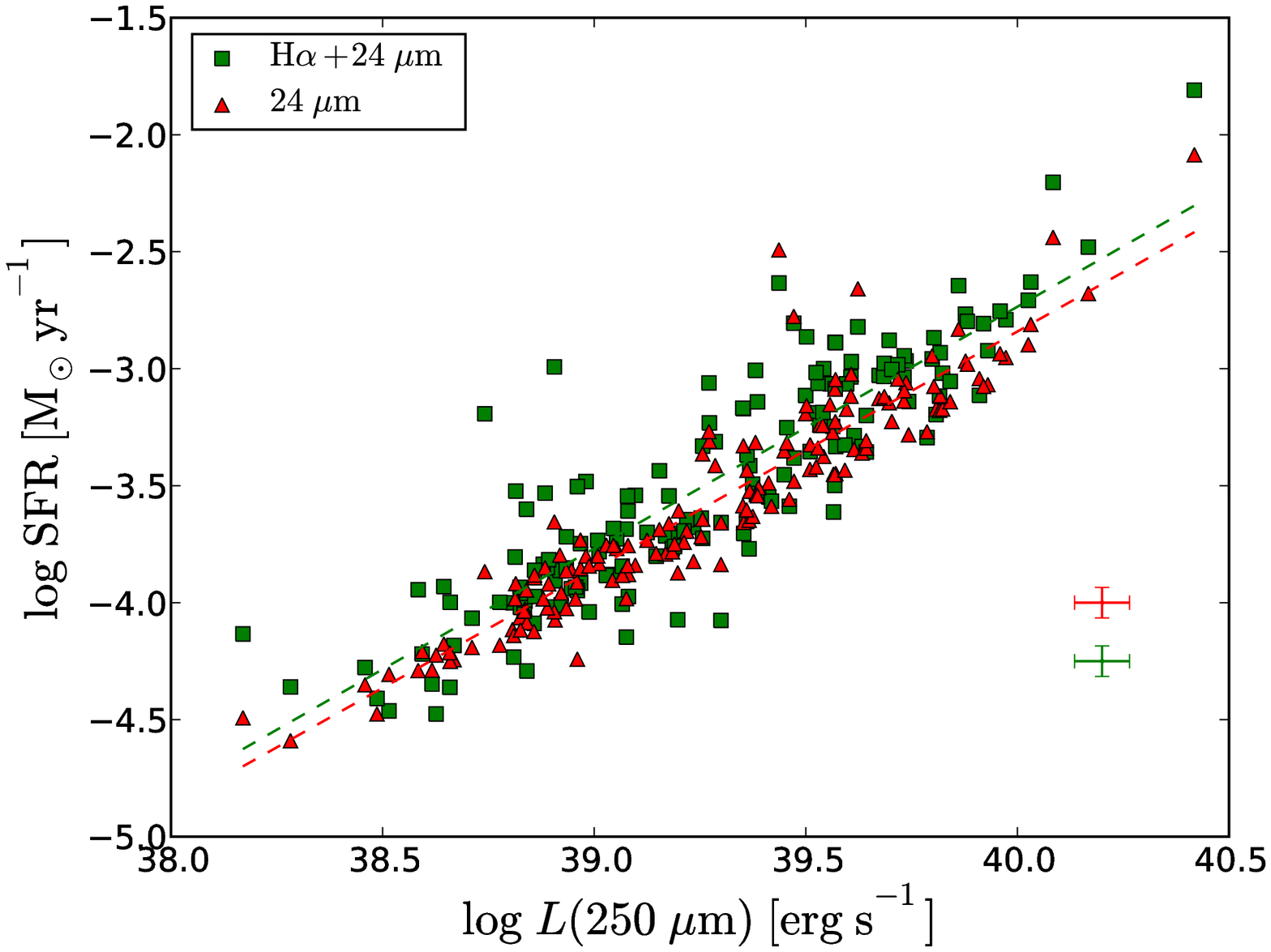}
\caption{{\it Upper panel:} H$\alpha$ and 24\,\micron\ emissions as a function of the 250\,\micron\ SPIRE emission for the 159 extracted sources. {\it Bottom panel:} Star formation rate from 24\,\micron\ and from H$\alpha$+24\,\micron\ emissions as a function of the 250\,\micron\ SPIRE emission. Typical uncertainties of 15\% are shown in the bottom right corner of each panel.}
\label{fig:spire250Ha24}
\end{figure}

To do so, we performed photometry on the H$\alpha$ and the {\it Spitzer} MIPS 24\,\micron\ images, using SExtractor to reproduce photometry on different images with the same centres and apertures. Thus, after having degraded (using a Gaussian function) the images to the 250\,\micron\ SPIRE resolution, we obtained the H$\alpha$ and 24\,\micron\ fluxes with the same input parameters as the ones used in the SPIRE 250\,\micron\ image. In the upper panel of Fig.~\ref{fig:spire250Ha24}, we show the H$\alpha$ and 24\,\micron\ luminosities as a function of the SPIRE 250\,\micron\ luminosity for the final 159 sources. Remarkable correlations appear between the H$\alpha$ and 24\,\micron\ luminosities and the 250\,\micron\ luminosity of these sources, confirming that the cool dust compact emission, as traced by the bright SPIRE 250\,\micron\ emission, is closely linked to SF. The 24\,\micron\ emission presents a very high Pearson correlation coefficient ($r_{24} = 0.94$, using the logarithmic values of the luminosities, as displayed in Fig.~\ref{fig:spire250Ha24}), while, although rather strongly correlated ($r_{\rm H\alpha} = 0.83$) with the 250\,\micron\ emission, the H$\alpha$ presents more scatter. That the 250\,\micron\ emission correlates better with the 24\,\micron\ emission is not surprising because both are linked to dust, either cool for 250\,\micron\ or warm for the 24\,\micron, while the H$\alpha$ emission directly reflects the ionising photons of hot OB stars. Part of the scatter of the H$\alpha$ observed luminosity can also be due to emission extincted by dust. A certain scatter between L(250\,\micron) and SFR is therefore expected because of variations in dust temperature, metallicity, maybe a varying initial mass function, and although minimised in our analysis, possibly heating due to the interstellar radiation field.

As the H$\alpha$ and 24\,\micron\ compact emissions are standard star formation rate (SFR) tracers, we can use the calibrations given by \citet{2007ApJ...666..870C,2010ApJ...714.1256C} for \HII\ regions to obtain the SFR of these sources. To calibrate the SFR from the 24\,\micron\ emission, we use Eq.~13 in \citet{2010ApJ...714.1256C} while for the combined H$\alpha$+24\,\micron\ SFR calibration, we use their Eq.~16. We can therefore obtain the calibration to recover the SFR from the 250\,\micron\ emission alone, by matching the existing H$\alpha$+24\,\micron\ and 24\,\micron\ SFR for \HII\ regions. 
To recover the H$\alpha$+24\,\micron\ SFR, our best fit leads to
\begin{equation} \label{eq:sfrMix}
{\rm SFR\ [M_\odot\, yr^{-1}]} = 8.71 \times 10^{-45} L(250\,\mu{\rm m})^{1.03} \quad ,
\end{equation}

\noindent where $L(250\,\mu{\rm m})$ is in erg~s$^{-1}$. To recover the SFR(24), we need the following calibration:
\begin{equation} \label{eq:sfr24}
{\rm SFR\ [M_\odot\, yr^{-1}]} = 3.47 \times 10^{-44} L(250\,\mu{\rm m})^{1.02} \quad ,
\end{equation}

\noindent also with $L(250\,\mu{\rm m})$ in erg~s$^{-1}$. The uncertainties are 0.04 and 0.03 for the exponents in Eqs.~\ref{eq:sfrMix} and \ref{eq:sfr24}, respectively, while the calibration constants have uncertainties of 4.0 and 2.7\%, respectively. The two equations show rather similar exponents, compatible to unity within $1 \sigma$, and reflect the almost linear behaviour of the compact 250\,\micron\ emission with respect to SFR. These recipes would then be valid to obtain SFR in \HII\ regions using the compact 250\,\micron\ emission for $10^{38} \leq L(250\,\mu{\rm m}) \leq 10^{40.5}$\,erg\,s$^{-1}$, approximately.

\vspace{-.5cm}
\section{SPIRE emission distributions for \HII\ regions} \label{sec:mor}

\begin{figure*}
\centering
\vspace{-1cm}
\includegraphics[width=0.32\textwidth]{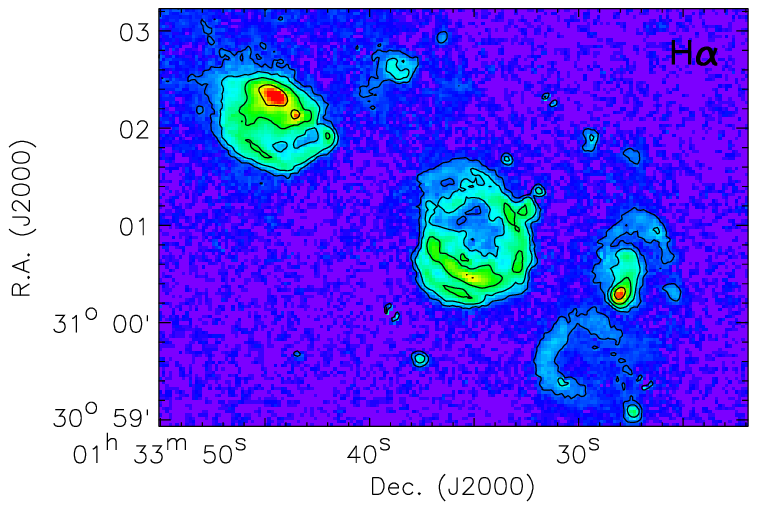}
\includegraphics[width=0.32\textwidth]{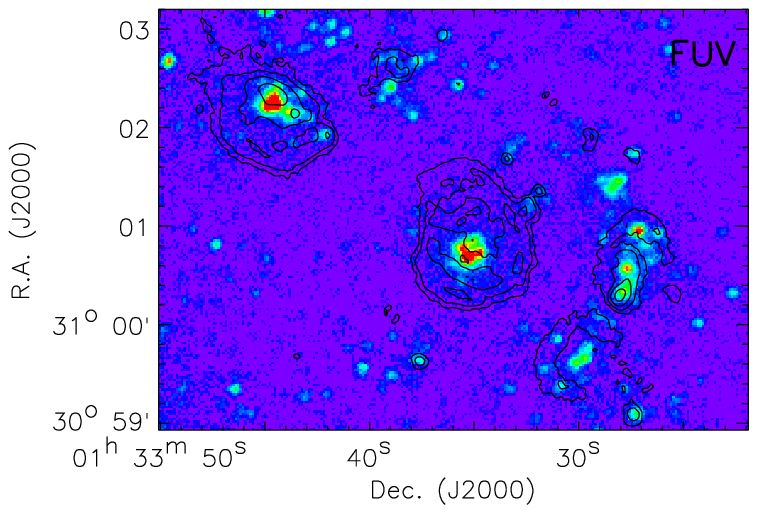}
\includegraphics[width=0.32\textwidth]{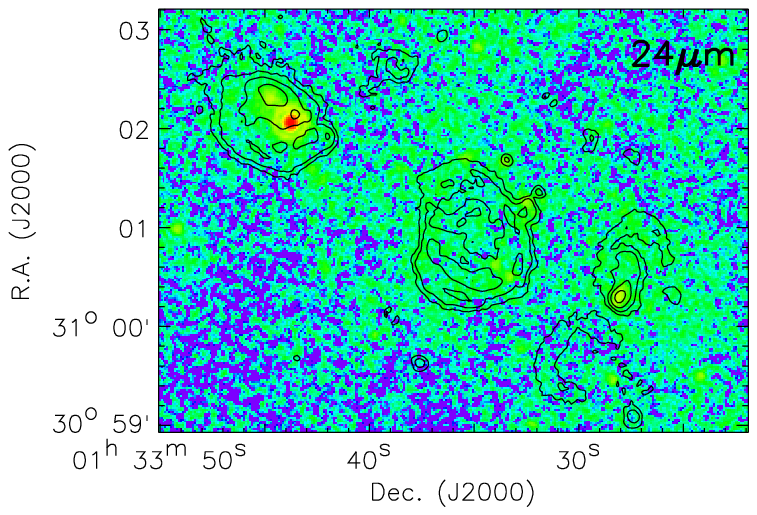}\\
\vspace{-1cm}
\includegraphics[width=0.32\textwidth]{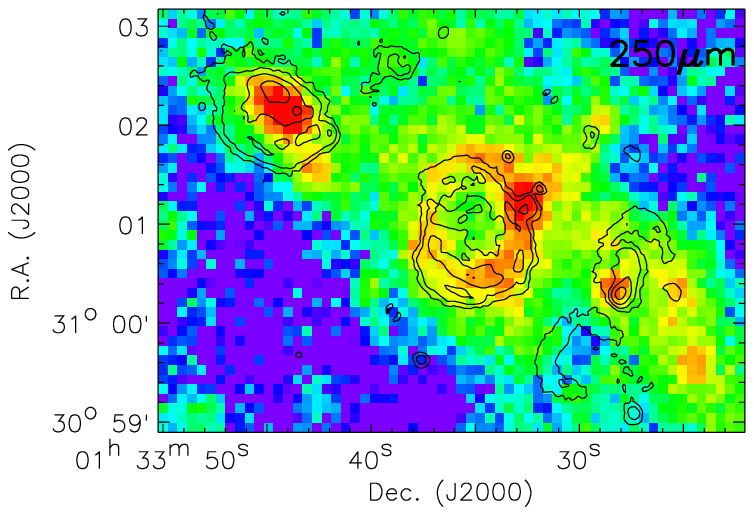}
\includegraphics[width=0.32\textwidth]{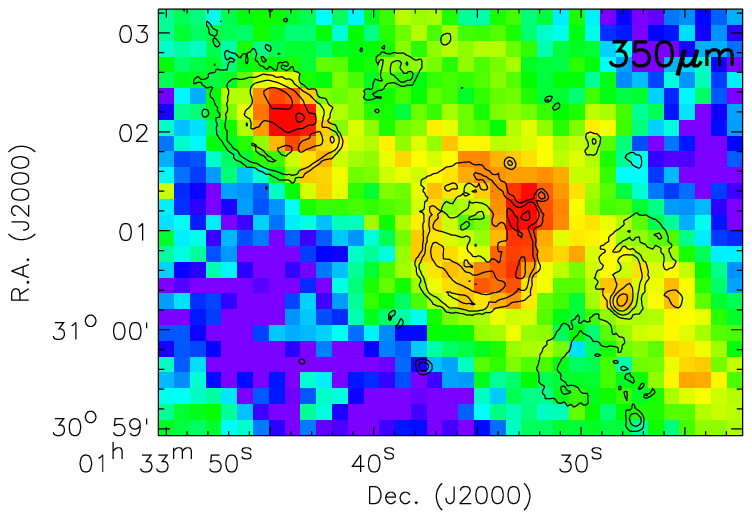}
\includegraphics[width=0.32\textwidth]{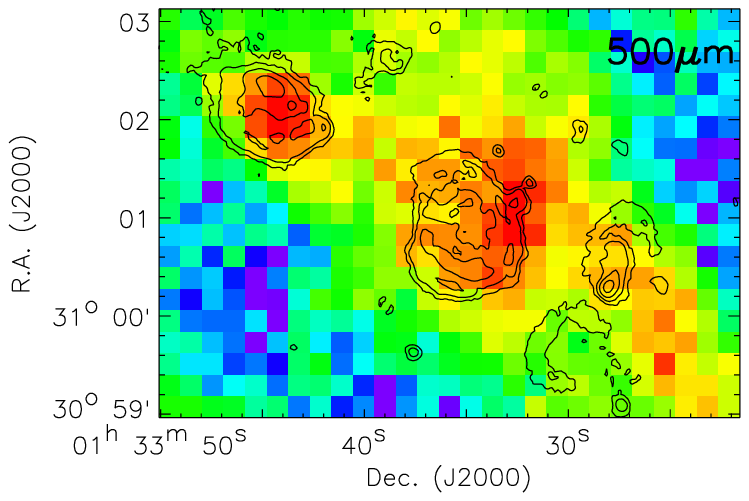}
\caption{Top-left: Continuum-subtracted H$\alpha$ image of a set of large shells in the outer north part of M\,33. Contours are overlaid in this image to better enhance the shell features (levels are at 20, 49, 122, 300 of H$\alpha$ emission measure) and repeated in the other images for comparison. From top to bottom and from left to right: FUV (res. 4\farcs2), 24\,\micron\ (res. 5\farcs7), 250\,\micron\ (res. 19\farcs7), 350\,\micron\ (res. 27\farcs0), and 500\,\micron\ (37\farcs2) images.}
\label{fig:shell}
\end{figure*}

The high resolution of the {\it Herschel} data for M\,33 allows us to perform a deeper comparison of the emission distributions within the star-forming regions. Following the work in \citet{2009ApJ...699.1125R} and \citet{2010MNRAS.402.1635R}, we focus on SPIRE emission in the interior of the most luminous \HII\ regions. {\it Herschel} wavelengths are particularly well-suited for this study because they trace the cool dust components that play a primordial role in \HII\ regions. 
From inspection of the two most luminous \HII\ regions in M\,33, NGC\,604 and NGC\,595, we see that in general the SPIRE-bands are displaced from the H$\alpha$ and FUV emission in the interior of the regions and have a more diffuse component extended towards the outer parts. Similar displacement between the infrared and FUV emissions has already been noticed by \citet{2005ApJ...633..871C} for about half of the studied \HII\ regions in M\,51a. 
As H$\alpha$ corresponds to the last $\sim10$\,Myr of SF and FUV to the last $\sim100$\,Myr, we can propose two possible explanations for the displacement: 1) as H$_2$ forms on the dust grains, then everything (dust and gas) collapses to form stars, and it takes between 10 and 100\,Myr for all the reservoir of dust in a given region to be consumed this way; 2) the UV field emitted mainly by OB stars and other non-ionising stars can be so strong that it pushes away the dust from the star-forming region in less than $\sim100$\,Myr. 

Support for the second explanation is the emission distribution for the most prominent shells in the north part of M\,33 presented in Fig.~\ref{fig:shell}. We compare in this figure the emissions at H$\alpha$, FUV, 24\,\micron, and SPIRE bands in a set of H$\alpha$-shells located in the north part of the galaxy. The location of these shells within the galaxy (white rectangle in Fig.~\ref{fig:spire250cat}) is in a region where no significant 8\,\micron\ or 24\,\micron\ emission has been detected, but where diffuse emission is observed in the SPIRE bands. 
The PACS bands (100\,\micron\ and 160\,\micron) were also examined, and we found some diffuse emission in the 160\,\micron\ but no significant emission in the 100\,\micron\ band. As shown in Fig.~\ref{fig:shell}, the shells have strong FUV emission in their centres. The northern shell seems to have a knot where the emissions at FUV and H$\alpha$ coincide, but the FUV is very nicely located in the centre for the bigger middle shell, suggesting that the shell could have been created by the stellar winds (SW) or supernova (SN) explosions coming from the stars in the cluster emitting at FUV. Using the H$\alpha$ image, the radius of the central H$\alpha$ shell is estimated to be $\sim$150\,pc. Assuming an expansion velocity of $\sim$50\,km\,s$^{-1}$ for the H$\alpha$ shell consistent with previous observations of H$\alpha$ expanding shells in \HII\ regions \citep{2005A&A...430..911R}, we derive a kinematical age for the central shell of $\sim$3\,Myr. Dynamical ages of a few Myr are consistent with ages estimated for shells in dwarf galaxies \citep{1998ApJ...506..222M}. 
It is remarkable that no emission at 24\,\micron\ is seen in the central and southern shells, but there is emission from cool dust in the three SPIRE bands. We suggest that, during the time the shell is forming, the dust is mixed with ionised gas inside the \HII\ region, a dust fraction is heated by the ionising radiation and emits at 24\,\micron, while the rest is cool and emits in the SPIRE bands. This could be the case for the northern shell in the process of formation. Later, when the shell is created by the SW and SNe, we observe FUV emission in the centre, H$\alpha$ emission from the shell, and cool dust outlining the H$\alpha$ structure (see the central shell where the 250\,\micron\ emission clearly delineates the H$\alpha$ shell). This would be the first evidence that the dust participates in the kinematics of the gas inside the \HII\ regions and that the impact of the SW and SNe into the ISM also affects the interstellar dust. The behaviour shown in Fig.~\ref{fig:shell} is also seen in a significant number of shells in the outer parts of the galaxy, and it would be very interesting to relate it with the existence of \HI\ bubbles observed in other galaxies \citep{2009ApJ...704.1538W}. A multi-wavelength comparison in a larger sample of shells would be needed to study this issue further.

\vspace{-.5cm}
\section{Conclusions}

For the Local Group spiral galaxy M\,33, using the unprecedented resolution and sensitivity of the {\it Herschel} SPIRE photometric data, we focus on the compact emission and conclude the following.

\begin{itemize}
\item We created a catalogue of 159 sources from the SPIRE 250\,\micron\ emission using a high detection threshold to focus on the compact emission of the cool dust; the compact emission sources associated to the cool dust follow the spiral pattern of the galaxy.
\item Very high Pearson correlation coefficients, $r_{24} = 0.94$ for the 24\,\micron\ emission and $r_{\rm H\alpha} = 0.83$ for the H$\alpha$ emission, confirm that the 159 sources are closely linked to SF.
\item By using standard (H$\alpha$+24\,\micron\ and 24\,\micron\ alone) SFR recipes for \HII\ regions provided by \citet{2007ApJ...666..870C,2010ApJ...714.1256C}, we are able to calibrate the 250\,\micron\ compact emission as an SFR tracer and provide conversion factors that can be used in further studies (Eqs.~\ref{eq:sfrMix} and \ref{eq:sfr24}).
\item The morphological study of a set of three H$\alpha$ shells shows that there is a displacement between far-ultraviolet and the SPIRE bands, while the H$\alpha$ structure is in general much more coincident with the cool dust.
\item The different locations of the H$\alpha$ and far-ultraviolet emissions with respect to the SPIRE cool dust emissions leads to a dynamical age of a few Myr for a set of H$\alpha$ shells and the associated cool dust.
\end{itemize}

\vspace{-.4cm}
\begin{acknowledgements} %%%%%%%%%%%%%%%%%%%%%%%%%%%%%%%%%%%%%%%%%%%%%%%%  
We thank R. Walterbos for the H$\alpha$ map. This work was partially supported by a Junta de Andaluc\' ia Grant FQM108, a Spanish MEC Grant AYA-2007-67625-C02-02, and a Juan de la Cierva fellowship. This research was supported by an MC-IEF within the 7$^{\rm th}$ European Community Framework Programme.
\end{acknowledgements}
\bibliographystyle{aa} %%%%%%%%%%%%%%%%%%%%%%%%%%%%%%%%%%%%%%%%%%%%%%%%%%   
\vspace{-.8cm}
\bibliography{herm33es}

\end{document}